\documentclass[preprint,12pt]{elsarticle}
\usepackage{amssymb}
\usepackage{amsmath}
\usepackage{url}
\usepackage{graphicx}
\usepackage{color} % added by boyukata

\def\be{\begin{equation}}
\def\ee{\end{equation}}
\def\bea{\begin{eqnarray}}
\def\eea{\end{eqnarray}}
\def\bdm{\begin{displaymath}}
\def\edm{\end{displaymath}}
\journal{Nuclear Physics A}

\begin{document}

\begin{frontmatter}

\title{Nuclear Structure Properties and Stellar Weak Rates for $^{76}$Se: Unblocking of the Gamow Teller Strength}

\author{Jameel-Un Nabi$^{1}$}
\author {Mavra Ishfaq$^{1}$}
\author{Mahmut B\"{o}y\"{u}kata$^{2}$}
\author {Muhammad Riaz$^{1}$}
\address{
$^1$Faculty of Engineering Sciences, GIK Institute of Engineering
Sciences and Technology, Topi 23640, Swabi, Khyber Pakhtunkhwa,
Pakistan}

\address{
$^2$Physics Department, Science and Arts Faculty, K\i r\i kkale
University, 71450, K\i r\i kkale, Turkey}

%\ead:$^a$ {\em jameel@giki.edu.pk}

\begin{abstract}

At finite temperatures ($\ge$ 10$^{7}$K), $^{76}$Se is abundant in
the core of massive stars and electron capture on $^{76}$Se has a
consequential role to play in the dynamics of core-collapse. The
present work may be classified into two main categories. In the
first phase we study the nuclear structure properties of $^{76}$Se
using the interacting boson model-1 (IBM-1). The IBM-1
investigations include the energy levels, $B(E2)$ values and the
prediction of the geometry. We performed the extended consistent-$Q$
formalism (ECQF) calculation and later the triaxial formalism
calculation (constructed by adding the \emph{cubic} term to the
ECQF). The geometry of $^{76}$Se can be envisioned within the
formalism of the potential energy surface based on the classical
limit of IBM-1 model. In the second phase, we reconfirm the
unblocking of the Gamow-Teller (GT) strength in $^{76}$Se (a test
case for nuclei having $N >$ 40 and $Z < 40$). Using the deformed
pn-QRPA model we calculate GT transitions, stellar electron capture
cross section (within the limit of low momentum transfer) and
stellar weak rates for $^{76}$Se. The distinguishing feature of our
calculation is a state-by-state evaluation of stellar weak rates in
a fully microscopic fashion. Results are compared with experimental
data and previous calculations. The calculated GT distribution
fulfills the Ikeda sum rule. Rates for $\beta$-delayed neutrons and
emission probabilities are also calculated. Our study suggests that
at high stellar temperatures and low densities, the
$\beta^{+}$-decay on $^{76}$Se should not be neglected and needs to
be taken into consideration along with electron capture rates for
simulation of presupernova evolution of massive stars.

\end{abstract}


\end{frontmatter}

\section{Introduction}

It is well known that electron captures on nuclei have a decisive
role during the end stages of presupernova and supernova phases of
massive stars \cite{bet79, bet90}. Capturing of electrons reduces
the degenerate relativistic electron gas pressure and ultimately
leads to the gravitational collapse of the core of a massive star.
The electron capture process also assists in reducing the
electron-to-baryon content, tilting the stellar matter towards more
neutron rich (hence heavy) nuclei. Some notable mentions of past
electron capture rate calculations include \cite{Niu11} (based on
relativistic energy density functionals), \cite{paa09, fan12} (based
on Skyrme finite temperature RPA), \cite{Moe97} (FRDM + RPA),
\cite{Hir93} (QRPA with separable interaction and deformed single
particle potential), \cite{Bor06} (continuum QRPA + density
functional theory) \cite{Mar99} (large scale shell model
calculation), \cite{Lan95} (shell model Monte Carlo method) and
\cite{Mut91} (shell model). See also Table 1 of Ref. \cite{Bor06}
for a list of global microscopic approaches to the calculation of
weak rates. In this paper we signify the transformation of a proton
into a neutron by GT$_{+}$ transitions.

Within the same major shell, the Independent Particle Model (IPM)
forbids the conversion of protons into neutrons for  $Z <$ 40 and
 $N > $ 40. This is due to the Pauli blocking of the
Gamow-Teller (GT) transitions. However it was debated that nuclear
correlation effects can make such transitions possible if nucleons
are allowed to move from one major shell ($fp$-shell) to another
($sdg$-shell). Also environments with finite temperature (as in core
of massive stars) can assist in doing the needful transformation. In
order to study the unblocking of GT strength with the help of finite
temperature and nuclear correlation effects, we select the nucleus
$^{76}$Se in this paper. This nucleus is also chosen because of its
copious production in the stellar core, during the presupernova and
supernova phases, of massive stars. The calculated GT strength
distribution is sensitive to nuclear structure properties, specially
the deformation parameter ($\beta_{2}$). In order to study the
nuclear structure properties of $^{76}$Se, we chose the interacting
boson model-1 (IBM-1) \cite{Iachello87}. The structure of $^{76}$Se
and its neighboring isotopes were previously studied within the
interacting boson model-2 (IBM-2)~\cite{Iachello87} where the
neutron and the proton bosons were separately taken into account
(e.g.~\cite{Kaup83,Radhi86,Speidel98,Turkan06,Boyukata08}). These
studies were mostly performed at low energy levels. Along $Z=32$
isotopic chain, the sub-shell effects were studied in
Ref.~\cite{Kaup83} where it was concluded that the sub-shell effect
occurs for $N\leq40$ in spite of the absence of this effect for
$N\geq42$. Radhi and Stewart focused on the mixed-symmetry states at
low-spin level scheme of $^{76}$Se~\cite{Radhi86} within the concept
of F-spin in the framework of IBM-2. To identify the effects of the
shell closure $N=50$, the $g$ factors of the $0^+_{\beta}$,
$2^+_{\gamma}$, $4^+_{gs}$ triplet states, appearing along the
stable $^{74-82}$Se isotopes, were experimentally measured and
compared with the IBM-2 results~\cite{Speidel98}. The energy levels,
B(E2) and B(M1) transition probabilities of this isotopic chain were
investigated in detail by using different sets of IBM-2 Hamiltonian
parameters in recent works~\cite{Turkan06,Boyukata08}.

The present work can be divided into two main phases. In the first
phase, because of the  behavior of $^{76}$Se, two kinds of IBM-1
calculation were performed to analyze it's structural properties.
This nucleus looks  spherical since the phonon multiplets picture
occurs as ($0_{gs}^+$), ($2_{gs}^+$), ($4_{gs}^+$, $2_{\gamma}^+$,
$0_{\beta}^+$), ($6_{gs}^+$, $3_{\gamma}^+$, $2_{\beta}^+$), ...  at
the low energy spectra. Despite this spherical behavior, its energy
ratio in the ground state band is $R_{4/2}=2.38$ (located in between
$R_{4/2}^{U(5)}=2.00$ and $R_{4/2}^{O(6)}=2.50$, which are the
typical ratios of the spherical and the $\gamma$-softness cases,
respectively). To move in between U(5) and O(6) symmetries, we chose
the extended consistent-$Q$ formalism (ECQF)~\cite{Lipas85} for the
$first$ calculation. We also performed the $second$ calculation by
adding the $cubic$ term~\cite{Heyde84} to the ECQF Hamiltonian to
study its effect on the energy levels in $\gamma$-band. The $B(E2)$
values were also calculated and compared with experimental data for
$^{76}$Se. After complete parametrization of the Hamiltonian, the
geometry of $^{76}$Se was predicted by plotting potential energy
surface as a function of deformation parameters ($\beta$,$\gamma$)
within the two formalisms of IBM-1. In the second phase, the IBM-1
(triaxial formalism calculation) energy levels were used as parent
excited states of $^{76}$Se to calculate stellar weak rates using
the the proton-neutron quasiparticle random phase approximation
(pn-QRPA) model \cite{Hal67}.

It is no surprise that different theoretical models have been used
in the past to calculate weak rates in stellar matter. There are two
major issues in calculation of stellar weak rates. There are
hundreds of nuclear species present in the stellar matter and
ideally one would like to calculate weak rates for all these nuclei.
The second problem is even more challenging. High temperatures
prevailing in the stellar matter populate parent excited states and
the total weak rate has a finite contribution from these excited
states. Most of the nuclear models would calculate ground state
GT$_{+}$ distributions (and maybe also for a few low-lying parent
states) and then adopt the so-called Brink-Axel hypothesis (BAH)
\cite{Bri55} for high-lying energy levels.  BAH states that the GT
resonance resides at the same relative energy in daughter for all
parent excited states akin to the ground state. Special mention of
weak rate calculations covering a wide range of nuclei would include
the IPM calculation \cite{ful82,ful82a}, the pn-QRPA
\cite{nab99,nab99a,nab04} calculation and the large scale shell
model (LSSM) \cite{lan00} calculation. The pn-QRPA model, when used
with a schematic separable interaction, has a distinguishing
feature. It allows a state-by-state calculation of ground state and
\textit{all} excited states GT distributions in a microscopic
fashion. In other words BAH is not assumed in the pn-QRPA
calculation (it is used in the IPM and LSSM calculation). The
reliability of pn-QRPA calculation hence increases many fold as
excited states contribute effectively to the total weak rate in the
core of massive stars. The pn-QRPA model also possesses the
additional advantage that it can be employed for any arbitrarily
heavy system of nucleons.

Previous works~\cite{Nabi16,Nabi17} where both IBM-1 and pn-QRPN
calculations were put into action  were performed for the waiting
point (WP) nuclei $^{60}$Zn, $^{64}$Ge, $^{68}$Se, $^{72}$Kr,
$^{76}$Sr, $^{80}$Zr, $^{84}$Mo, $^{88}$Ru, $^{92}$Pd and $^{96}$Cd
along $N=Z$ chain. These included the investigations of the nuclear
structure properties and their geometries. All these nuclei were
exotic (short live nuclei) and displayed varying behavior along the
line. B(E2:$2^+_{1}\rightarrow0^+_{1}$) was experimentally detected
for only two nuclei; $^{68}$Se and $^{72}$Kr. Unlike the past exotic
nuclei, $^{76}$Se, considered in this paper, is a stable nucleus and
has enough relevant experimental data. Further one of the main aims
of the current work is to study the unblocking of GT strength.

The Pauli blocking mechanism of allowed neutron hole orbits
available via the GT operator to protons capturing electrons in the
IPM was discussed in length by Fuller \cite{ful82a}.   It was
concluded that capture on free protons dominates the neutronization
rate in this region. It was later argued that Pauli blocking would
be overcome by finite temperature  and nuclear correlations effects
\cite{Lan03}.  GT transitions are thermally unblocked primarily as a
result of the excitation of neutrons from the $fp$ shell into the
$g_{9/2}$ orbital. This unblocking allows GT transitions within the
$fp$ shell, leading to domination of electron capture rates on
nuclei rather than free protons \cite{Hix03}, in contradiction to
previous findings. On the other hand, Cooperstein and Wambach
\cite{coo84} noted, from an investigation based on the random phase
approximation, that electron capture on neutron-rich nuclei with
protons in the $fp$ shell and  N $>$ 40 can compete with capture on
free protons if one considers forbidden transitions in addition to
allowed ones. The unblocking of GT strength in $^{76}$Se was
experimentally verified independently by (d, $^{2}$He)
charge-exchange reaction experiment at KV1 Groningen \cite{gre08}
and by the (n, p) reaction at TRIUMF \cite{hel97}. Different
theoretical models also confirmed this unblocking by calculating
GT$_{+}$ transitions for $^{76}$Se. Here we would like to mention
the deformed pn-QRPA calculation with density-dependent Skyrme
forces \cite{sar03} , LSSM \cite{zhi11} and deformed QRPA
calculation using a realistic two-body interaction \cite{ha15}.
Theoretical calculations on unblocking of GT strength are also
available in literature where other test cases were studied. In the
second phase of this paper, we use the pn-QRPA in a multi-shell
deformed single-particle space with a schematic separable
interaction to calculate GT$_{+}$ transitions in $^{76}$Se and
reconfirm the unblocking of GT strength for nuclei having $N >$ 40
and $Z < 40$. We later extend this model to calculate
weak-interaction mediated rates in stellar matter for $^{76}$Se.

Our paper is organized as follows. Section~2 briefly discusses the
necessary formalism of the IBM-1 and the pn-QRPA models used in our
calculation. Section~3 discusses and compares our calculation with
measured data and previous theoretical results. Section~4 finally
summarizes our findings.

\section{Formalism}

\subsection{Interacting Boson Model-1 (IBM-1)}
The IBM-1 model is one of the powerful approaches to study nuclear
structure properties of even-even nuclei. This model is a group
theoretical approach and is established on unitary algebra $U(6)$.
The $U(6)$ group has three possible sub-algebras denoted by $U(5)$,
$SU(3)$ and $O(6)$ known as \emph{Dynamical
Symmetries}~\cite{Iachello06}. In the IBM-1 model, different
versions of the Hamiltonian can be formulated depending on the
behavior of the given nucleus. The simple one is the consistent-$Q$
formalism (CQF)~\cite{Warner82, Warner83} including only two terms
$\hat Q\cdot\hat Q$ and $\hat L\cdot\hat L$ and the Hamiltonian can
be written as
\begin{equation}
\hat H_{CQF}= a_1 \,\hat L\cdot\hat L + a_2 \,\hat
Q\cdot\hat Q . \label{cqf}
\end{equation}
Here, $\hat L$ and $\hat Q$ are the angular momentum and quadrupole operators, respectively, as defined
by $\hat L=\sqrt{10}[d^\dag\times\tilde d]^{(0)}$,
$\hat Q=[d^\dag\times\tilde s+s^\dag\times\tilde d]^{(2)}+\overline{\chi}[d^\dag\times\tilde d]^{(2)}$.
The extended version of this simple Hamiltonian is called the
extended consistent-$Q$ formalism (ECQF)~\cite{Lipas85} and is
formed by adding an extra term $\hat n_d$ to the CQF Hamiltonian in
Eq.~(\ref{cqf}). The ECQF formalism provides the advantage to reach
all three symmetries and to move in between three dynamical
symmetries. The ECQF Hamiltonian~\cite{Lipas85} with three terms is
written as follows

\begin{equation}
\hat H_{ECQF}= \epsilon_d\,\hat n_d + a_1 \,\hat L\cdot\hat L + a_2
\,\hat Q\cdot\hat Q , \label{ham}
\end{equation}
where, $\hat n_d$ is the boson-number defined by $\hat
n_d=\sqrt{5}[d^\dag\times\tilde d]^{(0)}$. The constants
$\epsilon_d$,  $a_1$, $a_2$ are free parameters. These are fitted to
experimental data taken from National Nuclear Data Center
(NNDC)~\cite{NNDC16}. Hamiltonian~(\ref{ham}) includes $four$
parameters in total with $\overline{\chi}$ given in the quadrupole
operator $\hat Q$. In addition to the energy levels, the $B(E2)$
values can be calculated in IBM-1 model by using the $E2$ operator
$\hat T (E2)=e_b \hat Q$, where $e_b$ is the boson effective charge.

The triaxial formalism of IBM-1 was obtained by adding the $cubic$
interaction (also known as the $three-body$ term) to the given
Hamiltonian for triaxial effect in the IBM-1 as discussed in detail
in Refs.~\cite{Heyde84, Castanos84}. Some of the $O(6)$ like nuclei
were investigated within this formalism in Refs.~\cite{Casten85a,
Casten85b}. For the second calculation of the present application,
we formulate the triaxial case by adding this $cubic$ term to the
ECQF Hamiltonian~(\ref{ham}), like in~\cite{Heyde84}, as following
\begin{equation}
\hat H= \hat H_{ECQF} + \sum_{L} v_L [d^\dag d^\dag d^\dag]^{(L)}
\cdot [\tilde{d} \tilde{d} \tilde{d}]^{(L)}. \label{tri}
\end{equation}
The useful way to understand the effect of this $cubic$ term in the
$\gamma$~band is to look for the signature splitting of the
$\gamma$~band $S(J)$~\cite{Zamfir91}. One can test the
$\gamma$--~softness behavior within the given formula in
~\cite{Stefanescu07, Sorgunlu08} as
\begin{equation}
S(J) = \frac{E(J) - E(J-1)}{E(J) - E(J-2)} \cdot \frac{J(J+1) -
(J-1)(J-2)}{J(J+1) - J(J-1)}-1 , \label{sj}
\end{equation}
which describes whether the even--odd staggering appears in the
$\gamma$ band. In recent works \cite{Stefanescu07, Sorgunlu08},
transitional nuclei situated in between the spherical and
$\gamma$--~unstable cases were studied within the triaxial formalism
based on a Hamiltonian with three--body term. Fortunato $et.~
al.$~\cite{Fortunato11} added this term in the Hamiltonian of the
CQF to find the phase space of the triaxial region in between the
oblate and prolate shapes.

The other useful way for prediction of the nuclear shape is to plot
the potential energy surface $V(\beta,\gamma)$ obtained from the
IBM-1 Hamiltonian in the classical
limit~\cite{Dieperink80,Ginocchio80,Isacker81}. The energy surface
was found as in Ref.~\cite{Sorgunlu08}
\begin{equation}
V(\beta,\gamma) = E_0 + \sum_{n\geq1}\frac{N(N-1)\cdots(N-n+1)}{(1+\beta^2)^n} \sum_{kl} a^{(n)}_{kl}\beta^{2k+3l}\cos^l3\gamma,
\label{pes_climit}
\end{equation}
where the constant $E_0$ represents the binding energy of the core.
This energy surface can be written with the factors $a^{(n)}_{kl}$,
where $n$ = $1$, $2$, $3$ indicates  the order of the interaction in
the generators of U(6): \tiny{
\begin{eqnarray}
V(\beta,\gamma)&=&
\frac{N}{1+\beta^2}
\left(a^{(1)}_{00}+
a^{(1)}_{10}\beta^2\right)+
\nonumber\\&&
\frac{N(N-1)}{(1+\beta^2)^2}
\left(a^{(2)}_{00}+
a^{(2)}_{10}\beta^2+
a^{(2)}_{01}\beta^3\cos3\gamma+
a^{(2)}_{20}\beta^4\right)+
\nonumber\\&&\
\frac{N(N-1)(N-2)}{(1+\beta^2)^3}
\left(a^{(3)}_{00}+
a^{(3)}_{10}\beta^2+
a^{(3)}_{01}\beta^3\cos3\gamma+
a^{(3)}_{11}\beta^5\cos3\gamma+
a^{(3)}_{20}\beta^4+
a^{(3)}_{02}\beta^6\cos^23\gamma+
a^{(3)}_{30}\beta^6\right).
\label{pes_gen}
\end{eqnarray}
}\normalsize{The factors $a^{(n)}_{kl}$ ($n$ = $1$, $2$, $3$) are
also referred to as $one-body$, $two-body$, and $three-body$
interactions, respectively, as discussed in detail in
Ref.~\cite{Sorgunlu08}. Here, it can be reduced for
Hamiltonian~(\ref{ham}) as follows} \tiny{
\begin{eqnarray}
V(\beta,\gamma)&=&\epsilon N \frac{\beta^2}{1+\beta^2} + a_2 N
(N-1)\left[\frac{5+(1+\overline{\chi}^2)\beta^2}{(N-1)(1+\beta^2)} +
\frac{\left(\frac{2\overline{\chi}^2\beta^4}{7}-4\sqrt{\frac{2}{7}}\overline{\chi}\beta^3\cos(3\gamma)+4\beta^2\right)}{(1+\beta^2)^2}\right],
\label{pes_ecqf}
\end{eqnarray}
}\normalsize{which includes common free parameters used as constant
in the $ECQF$ formalism. By adding the $cubic$ interaction, we can
obtain the energy surface for triaxial formalism as} \tiny{
\begin{eqnarray}
V(\beta,\gamma)&=&\epsilon N \frac{\beta^2}{1+\beta^2} +
\nonumber\\&&
a_2 N (N-1)\left[\frac{5+(1+\overline{\chi}^2)\beta^2}{(N-1)(1+\beta^2)}
+(N-1) \frac{\left(\frac{2\overline{\chi}^2\beta^4}{7}-4\sqrt{\frac{2}{7}}\overline{\chi}\beta^3\cos(3\gamma)+4\beta^2\right)}{(1+\beta^2)^2}\right]+
\nonumber\\&&
v_3 N(N-1)(N-2) \frac{1}{30(1+\beta^2)^3} \sin^23\gamma,
\label{pes_cubic}
\end{eqnarray}
}\normalsize{where the $cubic$ interaction with $L=3$ is symbolized
by $v_3$ and it is commensurate to $\sin^23\gamma ~( = 1 - \cos^23\gamma )$.}

Both formalisms of the energy surface given in
Eqs.~(\ref{pes_ecqf})~ and ~(\ref{pes_cubic})~ include the shape
variables ($\beta$, $\gamma$), also known as deformation parameters.
These variables have similar role as in the collective model of Bohr
and Mottelson~\cite{Bohr98}. The role of these collective
deformations is that $\beta$ measures the axial deviation from
sphericity and the angle $\gamma$ controls the departure from axial
deformation. Within the energy surface of IBM-1, one can move in
between the three dynamical symmetries relating to the geometry of
nuclei. The $U(5)$, $SU(3)$ and $O(6)$ symmetries correspond to the
spherical, the axially deformed and $\gamma$-unstable shapes,
respectively.

\subsection{The pn-QRPA formalism for calculation of stellar weak rates}
The Hamiltonian of the pn-QRPA model was chosen as
\begin{equation}
H^{QRPA} = H^{sp} + V^{pair} + V ^{ph}_{GT} + V^{pp}_{GT}.
\label{Eqt. sp}
\end{equation}
Wave functions and single particle energies were calculated using
the Nilsson model (axially deformed). Pairing in nuclei was treated
within the BCS approximation. The proton-neutron residual
interaction occurred through particle-particle ($pp$) and
particle-hole ($ph$) channels in our model. In the pn-QRPA model
these interaction (force) terms were given a separable form as
$V_{GT}^{ph}$ for the particle-hole Gamow-Teller force  and
$V_{GT}^{pp}$ for the particle-particle Gamow-Teller force. The
other parameters for calculation of weak rates are the Nilsson
potential parameters, the pairing gaps, the nuclear quadrupole
deformation, and the Q-value of the reaction. Nilsson-potential
parameters were adopted from Ref. \citep{Nil55} and the Nilsson
oscillator constant was chosen as $\hbar \omega=41A^{-1/3}(MeV)$
(the same for protons and neutrons). Pairing gaps used in the
present work were taken as $\Delta _{p} =\Delta _{n} =12/\sqrt{A}
(MeV)$. Nuclear quadrupole deformation parameter selection would be
discussed in Section 3. Q-values were taken from the recent mass
compilation of Audi and collaborators \citep{aud12}.

Calculation of stellar weak rates using the pn-QRPA  model was first
performed by Nabi et al. \cite{nab99}. The decay rate from parent
state ($\mathit{i}$th) to the daughter state ($\mathit{j}$th) of the
nucleus is given by

\begin{equation}
\lambda_{ij}^{ec(pd)} =ln2
\frac{f_{ij}^{ec(pd)}}{(ft)_{ij}^{ec(pd)}}. \label{d_rate}
\end{equation}
The $f_{ij}^{ec(pd)}$ are the phase space integrals for electron
capture (positron decay) reaction and are functions of stellar
temperature ($T$), density ($\rho$) and Fermi energy ($E_{f}$) of
the leptons. They are explicitly given by
\begin{equation}
f_{ij}^{ec} \, =\, \int _{w_{l} }^{\infty }w\sqrt{w^{2} -1}
 (w_{m} \, +\, w)^{2} F(+Z,w)G_{-} dw.
 \label{ec}
\end{equation}
and by
\begin{equation}
f_{ij}^{pd} \, =\, \int _{1 }^{w_{m}}w\sqrt{w^{2} -1} (w_{m} \,
 -\, w)^{3} F(- Z,w)(1- G_{+}) dw,
 \label{pd}
\end{equation}
In Eqs. ~(\ref{ec}) and ~(\ref{pd}), $w$ is the total energy of the
electron including its rest mass. $w_{m}$ is the total $\beta$-decay
energy,
\begin{equation}
w_{m} = m_{p}-m_{d}+E_{i}-E_{j},
\end{equation}
where $m_{p}$ and $E_{i}$ are masses and excitation energies of the
parent nucleus, and $m_{d}$ and $E_{j}$ of the daughter nucleus,
respectively. F($ \pm$ Z,w) are the Fermi functions and were
calculated according to the procedure adopted by Gove and Martin
\cite{Gov71}. G$_{\pm}$ are the Fermi-Dirac distribution functions
for positrons (electrons).

The $ft$ values are associated with the reduced transition
probability of nuclear transitions B$_{ij}$,

\begin{equation}\label{ft} (ft)_{ij}^{ec(pd)}=F/B_{ij}.
\end{equation}

The $F$ in Eq.~\ref{ft} represents a physical constant,
\begin{equation}
F=\frac{2ln2\hbar^{7}\pi^{3}}{g_{V}^{2}m_{e}^{5}c^{4}},
\end{equation}
with a value of 6146 $\pm$ 6s \cite{jok02}, and

\begin{equation}
B_{ij}=B(F)_{ij}+(g_{A}/g_{V})^2 B(GT)_{ij},
\end{equation}

\begin{equation}
B(F)_{ij} = \frac{1}{2J_{i}+1} \mid<j \parallel \sum_{k}t_{+}^{k}
\parallel i> \mid ^{2},
\end{equation}

\begin{equation}\label{gt}
B(GT)_{ij} = \frac{1}{2J_{i}+1} \mid <j
\parallel \sum_{k}t_{+}^{k}\vec{\sigma}^{k} \parallel i> \mid ^{2}.
\end{equation}
The value of the ratio of axial to vector coupling constant
($g_{A}/g_{V}$) was taken as~-~1.2694 \cite{nak10}. In Eq.~\ref{gt},
$t_{+}^{k}$ shows isospin raising operator while $\vec{\sigma}^{k}$
represents the spin operator.

The total electron capture/$\beta^{+}$ decay rate per unit time per
nucleus is finally given by
\begin{equation}
\lambda^{ec(pd)} =\sum _{ij}P_{i} \lambda _{ij}^{ec(pd)}.
\label{total rate}
\end{equation}
The summation was performed until satisfactory convergence was
achieved in our rate calculation.  The occupation probability of a
given parent excited state, P$_{i} $, was calculated assuming the
Boltzmann distribution.

In our calculation it was further assumed  that daughter excited
states, having energy larger than separation energy for neutrons
(S$_{n}$), would decay by emission of neutrons. From the daughter
nucleus the rate of neutron energy was calculated using the relation
\begin{equation}\label{ln}
\lambda^{n} = \sum_{ij}P_{i}\lambda_{ij}(E_{j}-S_{n}),
\end{equation}
for all $E_{j} > S_{n}$.

The $\beta$-delayed neutron emission probability was calculated
using
\begin{equation}\label{pn}
P^{n} =
\frac{\sum_{ij\prime}P_{i}\lambda_{ij\prime}}{\sum_{ij}P_{i}\lambda_{ij}},
\end{equation}
here j$\prime$ are the daughter states for which E$_{j\prime}>
S_{n}$. In Eqs.~(\ref{ln} and \ref{pn}), $\lambda_{ij(\prime)}$, for
the transition $i$ $\rightarrow$ $j(j\prime)$, is sum of electron
capture and positron decay rates.

The Q-values, adopted in our calculation, were taken from the recent
mass compilation of Audi et al. \cite{aud12}.

The pn-QRPA results for calculated GT strength was quenched by a
factor of $f_{q}^{2}$ = (0.55)$^{2}$ \cite{nab15b}. The
re-normalized Ikeda sum rule translates to

\begin{equation}
(ISR)_{renorm} = \sum B(GT)_{-}- \sum B(GT)_{+}\cong
3f_{q}^{2}(N-Z). \label{Eqt. ISR}
\end{equation}
The difference in our calculated  (and quenched) strength values was
7.14. This was very close to the re-normalized Ikeda sum rule value
of 7.26. The 98.34\% fulfillment of Ikeda sum rule deserves a
special mention in this work.

For details of the pn-QRPA model Hamiltonian and solution of the
QRPA equation with separable GT forces we refer to \cite{mut92}.

\subsection{Electron capture cross sections}

The nucleus $(N, Z)$ captures electron of incident energy (E$_e$)
and decay weakly as
\begin{equation}
(N, Z) + e^- \longrightarrow (N+1, Z-1)^* + {\nu}_{e}
\label{decay}
\end{equation}
The incident electron energy maybe distributed into two parts, a
part of it is absorbed by daughter nucleus $(N+1, Z-1)$ to change
its state from initial E$_{i}$ to final  E$_{f}$ state and the
remaining energy is carried out by neutrino $\nu_e$. Using energy
conservation in Eq.~(\ref{decay}), the energy of the outgoing
neutrino $E_{\nu}$ may be calculated using
\begin{equation}\label{nu}
E_{\nu}=E_{e}-Q+E_{i}-E_{f},
\end{equation}
where the $Q$ value is simply the difference of the measured masses
of parent and daughter nuclei. The nuclear reaction cross section
calculation for the reaction (\ref{decay}) is governed by the weak
Hamiltonian
\begin{equation}
\widehat{H}=\frac{G}{\sqrt2}j_\mu^{lept}\widehat{J}^{\mu},
\end{equation}
where  G=G$_F$ cos$\theta_c$, G$_F$ is the Fermi coupling constant,
$\theta_c$ is Cabibbo angle, j$_\mu^{lept}$ and $\widehat{J}^{\mu}$
are the leptonic and hadronic currents, respectively (for further
details, see e.g. \cite{div13}).

We used the limiting condition q$ \longrightarrow 0$ of low momentum
transfer. Using this assumption the impact on the total electron
capture cross section is provided only by the GT transitions
\cite{Goo80}. The pn-QRPA calculated energy eigenvalues of the
parent nucleus $|i\rangle$ were replaced manually  by those
calculated using the IBM-1 calculation (triaxial formalism). The
excited states of daughter nucleus $|f\rangle$ and GT transitions
were calculated using the pn-QRPA equations with separable GT forces
\cite{mut92}. The total stellar EC cross section, as a function of
incident electron energy and stellar temperature, is given by
\begin{equation}
\begin{split}
\sigma(E_e,T)=\frac{G_F^{2}cos^2\theta_c}{2\pi}\sum\limits_{i}F(Z,E_e)\frac{(2J_{i}+1)\exp{(-E_i/kT)}}{G(A,Z,T)}\\
\times \sum\limits_{J,f}(E_e-Q+E_i-E_f)^{2}\frac{|\langle
i|\sigma\tau_{+}|f\rangle|^{2}}{(2J_{i}+1)}.
\label{ECcs}
\end{split}
\end{equation}
In Eq.~(\ref{ECcs}), F(Z, E$_e)$ is the well known Fermi function
and was calculated according to the prescription given in Ref
\cite{Gov71}. G(A, Z, T) are the nuclear partition functions and
were calculated using the recipe of \cite{Nab16}.

\section{Results and Discussions}
The energy spectra of $^{76}$Se were calculated by fitting the
Hamiltonian parameters in Eqs.~(\ref{ham}) and~(\ref{tri}). This
spectra includes $17$ experimentally known levels in the ground
state ($g.s.$), $\gamma$ and $\beta$ bands~\cite{NNDC16}. In
addition to the calculation of the known levels, the prediction was
also made for unknowns in the $\gamma$ and $\beta$ bands. For this
application, we performed two calculations; the \emph{first} one is
denoted by IBM-1 for the EQCF formalism and the \emph{second} by
IBM-1$^{*}$ with $star$ for the formalism in Eq.~(\ref{tri})
including \emph{cubic} term.

\begin{figure}
\begin{center}
\includegraphics[width=13.5cm]{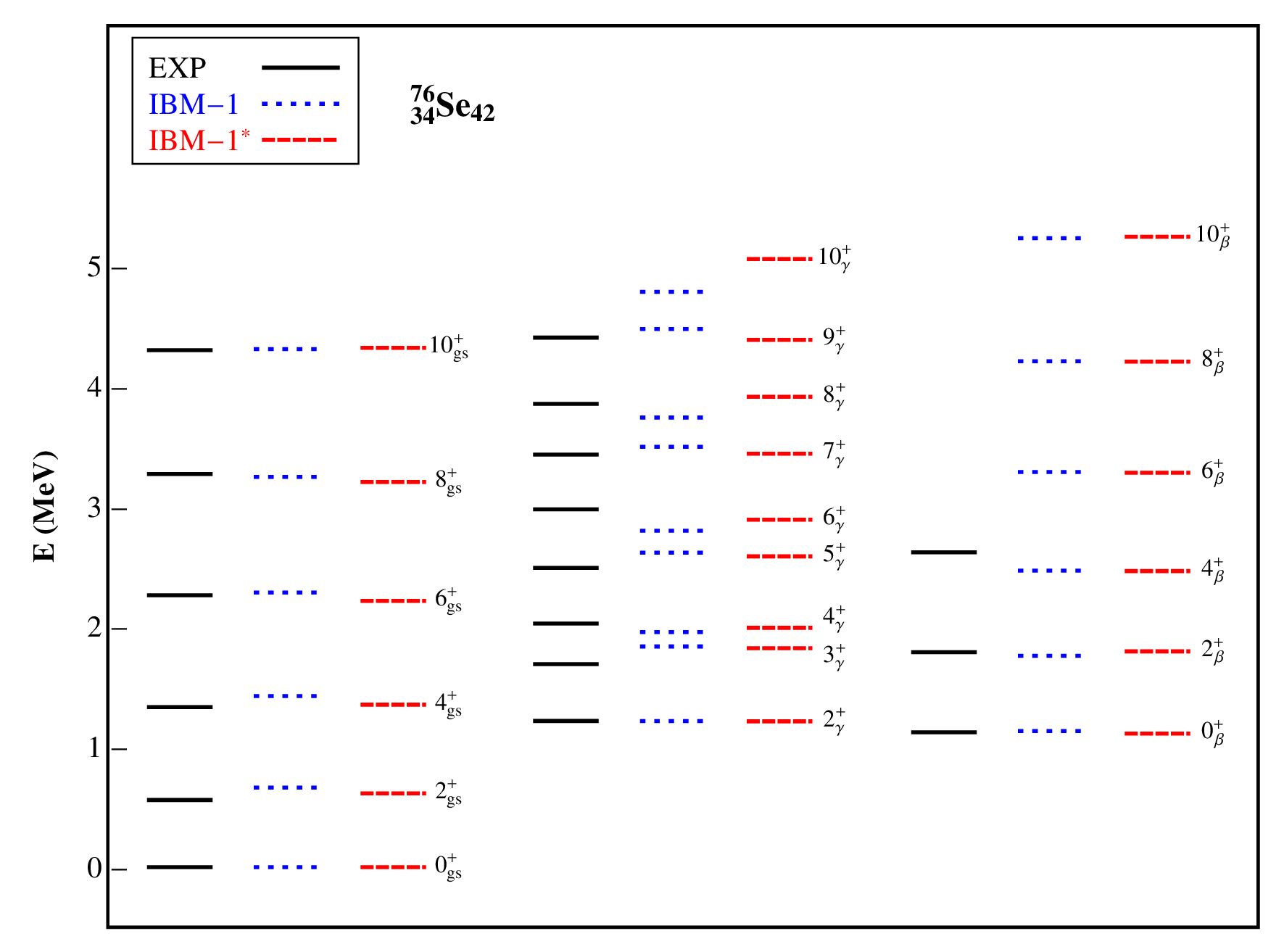}
\end{center}
\caption{(Color online) The measured ~\cite{NNDC16} and calculated
levels for ground-state (gs), $\gamma$ and $\beta$ bands in the
energy spectra of $^{76}$Se. The bold lines indicate experimental,
the dotted and dashed lines denote the calculation of IBM-1 (without
$cubic$ term) and IBM-1$^{*}$ (with $cubic$ term), respectively.}
\label{f_en}
\end{figure}

For the fitting procedure of the \emph{first} calculation, we first
determined $\epsilon_d$ while keeping other constants as $zero$.
Later we fitted the other two parameters ($a_2$, $a_1$) by
minimizing the root-mean-square ($rms$) deviation. Finally,
$\overline{\chi}$ was determined by changing its value from $0$ to
${-\sqrt{7}}/2$, step by step, and for each step $\epsilon_d$,
$a_2$, $a_1$ were re-fitted to find minimum $rms$ value (see
Table~\ref{par}). The calculated energy levels (dotted lines) with
IBM-1 (EQCF) and the experimental data (bold lines) are illustrated
in Fig.~\ref{f_en}. As seen in this figure, the IBM-1 results are in
good agreement especially for the levels in $g.s.$ and $\beta$~bands
except for the levels in $\gamma$~band. An important feature of
abnormal levels called "staggering" appears in $\gamma$~band. The
calculated $\gamma$~band levels tend to be coupled as
($2_{\gamma}^+$), ($3_{\gamma}^+$, $4_{\gamma}^+$), ($3_{\gamma}^+$,
$4_{\gamma}^+$), ($5_{\gamma}^+$, $6_{\gamma}^+$),~...~; a typical
band structure of the $\gamma$-softness~\cite{Casten90}. However,
these couplings are not so close for experimental data and it is
difficult to predict whether $^{76}$Se exhibit exact
$\gamma$-softness picture. Nevertheless, this result gives a choice
to test the effect of \emph{cubic} term on the $\gamma$~band levels.
Later we added the \emph{cubic} term to the Hamiltonian~(\ref{ham})
for the \emph{second} calculation shown as IBM-1$^{*}$. We carried
out a similar fitting procedure as for the \emph{first} calculation
to determine the parameters of the Hamiltonian~(\ref{tri}). As a
result, we obtained a smaller $rms$ value as given in
Table~\ref{par} and this suggests that the $cubic$ term has a
positive effect on the $\gamma$~band levels. The \emph{second}
calculated levels (dashed lines) of $g.s.$ and $\beta$~bands are the
same as the \emph{first} one. The $\gamma$~band levels appear better
in the triaxial formalism since the odd-even couplings are separated
as seen in Fig.~\ref{f_en}.

\begin{figure}
\begin{center}
\includegraphics[width=13.5cm]{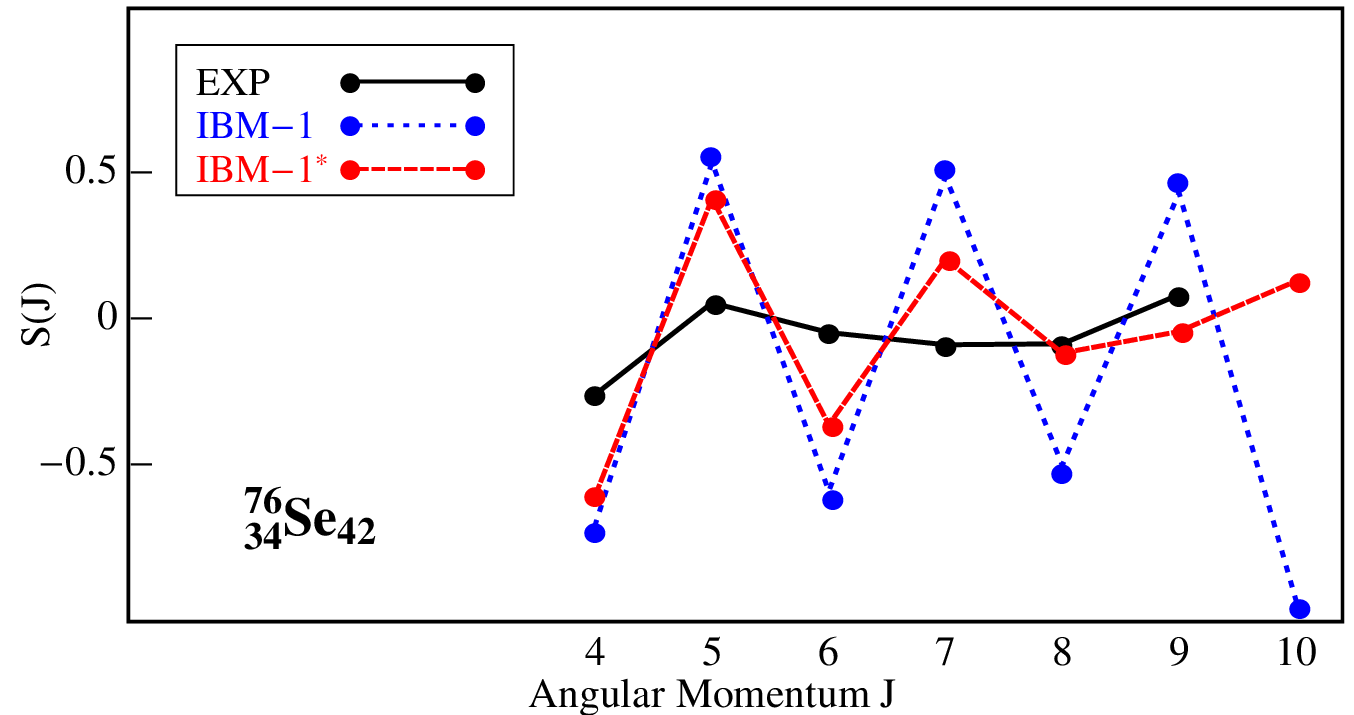}
\end{center}
\caption{(Color online) The experimental (bold line) and calculated
(dotted and dashed lines) odd-even staggering plots of $\gamma$
bands in $^{76}$Se. The exact data points of the staggering are
indicated by circles; the black, blue and red ones denote the
results of the experimental, IBM-1 (without $cubic$ term) and
IBM-1$^{*}$ (with $cubic$ term), respectively.} \label{f_sj}
\end{figure}

Another useful way is to look for the signature splitting function
$S(J)$, given in Eq.~(\ref{sj}), to study the effect of the $cubic$
term on the odd-even staggering in the $\gamma$~band. The $S(J)$ is
plotted as a function of the angular momentum-$J$ as shown in
Fig.~\ref{f_sj}. It is clearly seen that the sliding towards the
experimental data appears when the cubic term is inserted into the
ECQF Hamiltonian~(\ref{ham}). Especially, the high levels
$8_{\gamma}^+$, $9_{\gamma}^+$ exactly coincide with the
experimental points and this suggests that $cubic$ term improves
IBM-1$^{*}$ results.

We calculated the $B(E2)$ values by using boson effective charge
$e_{\rm b} = 0.097~eb$, fitted for the nuclei at $A\sim100$
region~\cite{Boyukata10}. As seen in Table~\ref{be2}, both results
of IBM-1 and IBM~-1$^{*}$ calculations are in good agreement with
the experimental data~\cite{NNDC16}. The $cubic$ term has also a
positive influence on the B(E2) values.

\begin{figure}
\begin{center}
\includegraphics[width=16cm]{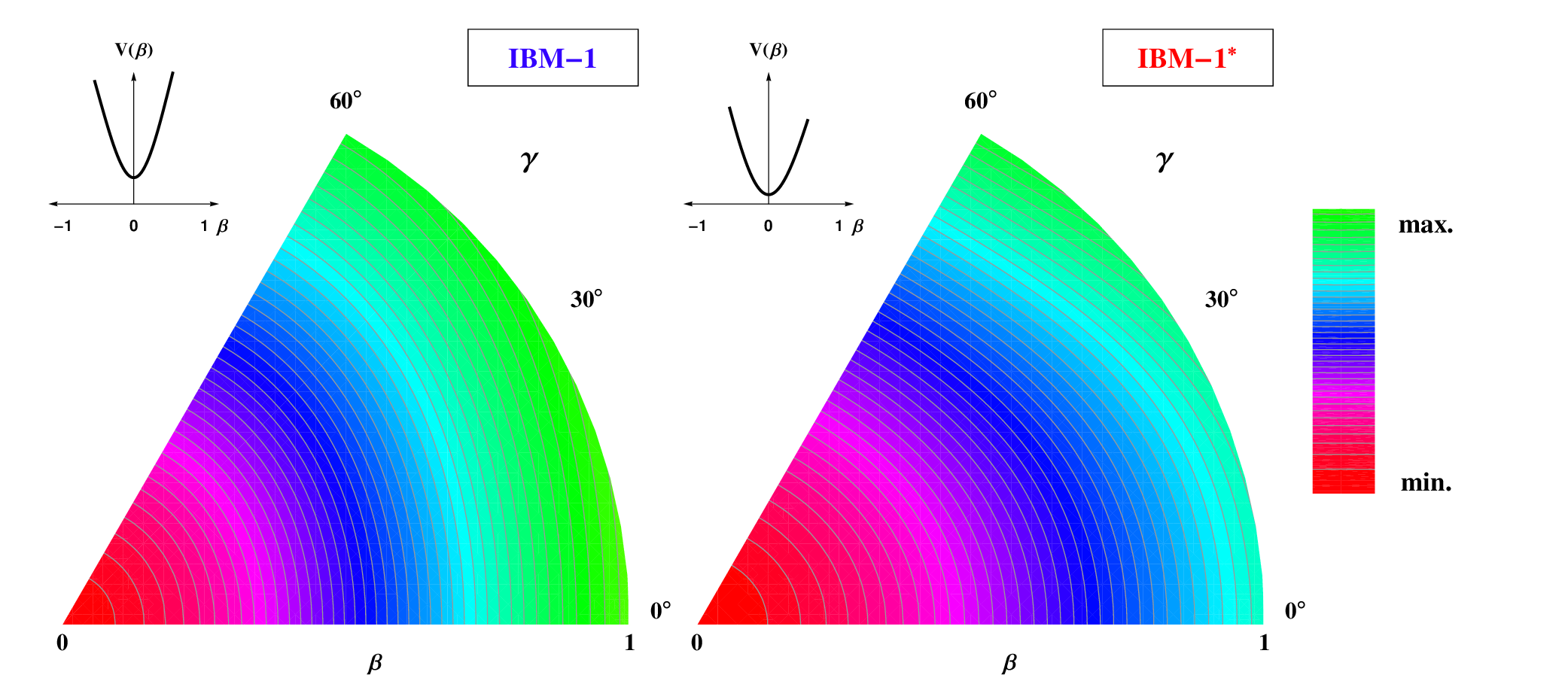}
\end{center}
\caption{(Color online) Potential energy surfaces as function of
$\beta$ and $\gamma$ for $^{76}$Se nucleus and also as function of
$\beta$ for $\gamma$ = $0$.} \label{f_pes}
\end{figure}

In addition to the energy levels and $B(E2)$ values of $^{76}$Se, we
predicted its geometry by plotting the potential energy surface as
function of deformation parameters within both formalisms in
Eqs.~(\ref{pes_ecqf})~and~(\ref{pes_cubic}). The counterplots as a
function of $\beta$ and $\gamma$ are depicted in Fig.~\ref{f_pes}
along with its scale. Shown also is the energy surface as a function
of $\beta$ for $\gamma =0^\circ$. Both plots predict the given
nucleus to be spherical (using both formalisms of IBM-1 and
IBM-1$^{*}$). The effect of the $cubic$ term is not so evident in
this case, only the \emph{second} minimum is flatter than the
\emph{first} albeit not so much.

The calculated zero value of deformation parameter using the IBM
model is in contrast to the oblate deformation calculated using the
RMF model \cite{lal99}. The QRPA calculation of GT strength
distribution is a sensitive function of the deformation parameter
\cite{ha15}. Table~\ref{dv} shows the various values of $\beta_{2}$
for $^{76}$Se. At the end we decided to use the $\beta_{2}$ value
extracted from E2 transition \cite{ram01} in our pn-QRPA
calculation. The chosen value of $\beta_{2}$ was expected to give
better results as it was extracted from measured data.

\begin{figure}
\includegraphics[scale=0.5]{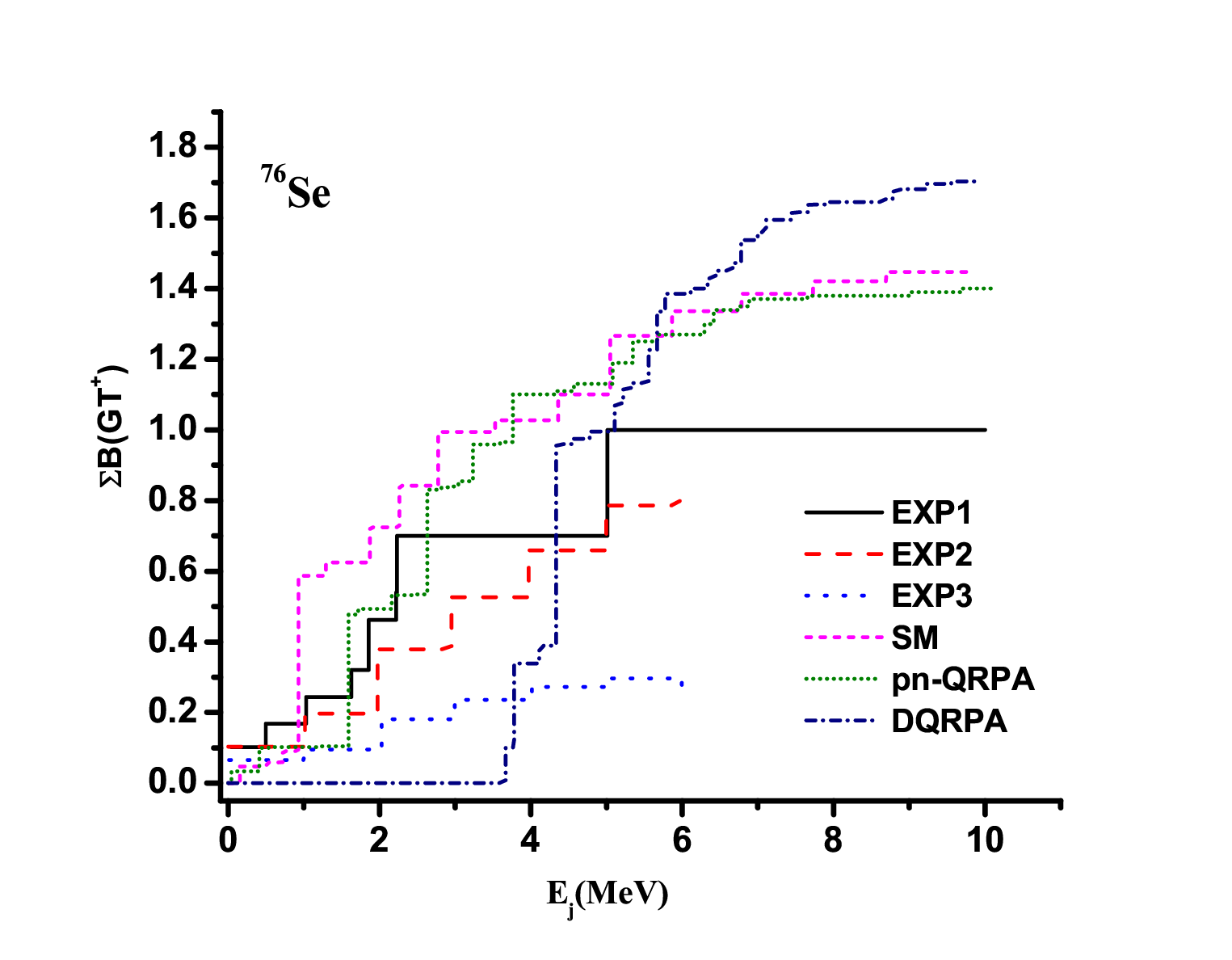}
\normalsize \caption {(Color online) Comparison of pn-QRPA
calculated GT strength for $^{76}$Se with experimental, shell model
\cite{zhi11} and DQRPA model \cite{ha15} results. EXP1 shows
measured values by Grewe et al. \cite{gre08} while EXP2 and EXP3 by
Helmer et al. \cite{hel97}.}\label{bgt}
\end{figure}

Fig.~\ref{bgt} shows cumulative GT$_{+}$ strength for $^{76}$Se.
Daughter excitation energy in units of MeV is represented on the
abscissa. The three calculated GT$_{+}$ strength distributions
include our, shell model \cite{zhi11} and the DQRPA \cite{ha15}
results. Shown also is the $^{76}$Se(d, $^{2}$He)$^{76}$As
experimental data of \cite{gre08} and the $^{76}$Se(n, p)$^{76}$As
data of \cite{hel97}. The (n, p) experiment performed by Helmer and
collaborators resulted in two sets of data. One using the multipole
decomposition analysis and the other using background subtraction
technique. These are represented by EXP2 and EXP3 in Fig.~\ref{bgt},
respectively. The figure shows very good comparison of our
calculation and shell model result. The DQRPA model calculation
results in largest calculated value of GT strength in $^{76}$Se.
There exists a considerable difference between pn-QRPA and DQRPA
calculated strength distributions. We attribute this diversity to a
different choice of single particle basis and interaction strength
parameters. Ha $\&$ Cheoun employed a deformed, axially symmetric
Woods-Saxon potential whereas we took a deformed Nilsson single
particle basis. They used the Brueckner G-matrix based on the CD
Bonn potential and used particle-particle and particle-hole strength
parameters as $g_{pp}$ = 0.99 and $g_{ph}$ = 1.15. Our corresponding
values were 0.04 and 0.35, respectively. Pairing gap parameter value
was taken as 1.7 by Ha $\&$ Cheoun to be compared with our chosen
value of 1.38. The deformation parameter $\beta_2$ adopted by Ha
$\&$ Cheoun was also slightly different. They took a value of 0.3 to
be compared with our value of 0.309. Our result is in decent
comparison with the (d, $^{2}$He) data. Fig.~\ref{bgt} is a clear
evidence of the unblocking of the GT strength. All the six
measurements and calculations support the unblocking of GT strength
in $^{76}$Se under terrestrial conditions caused due to nuclear
correlation effects. The GT strength may also be unblocked due to
finite temperature effects which we discuss later.

Table~\ref{centroid} shows the statistics of the calculated and
measured GT$_{+}$ strength distributions. Here it is noted that our
model places the GT centroid at  higher energy in daughter compared
with the shell model calculation. The shell model calculates a
slightly bigger total strength than our pn-QRPA model. It is  noted
that the total strength and centroid placement of the pn-QRPA
calculation is in decent agreement with the measured data. It is
pertinent to mention that no quenching factor was employed in shell
model and DQRPA calculations.

\begin{figure}
\includegraphics[scale=0.5]{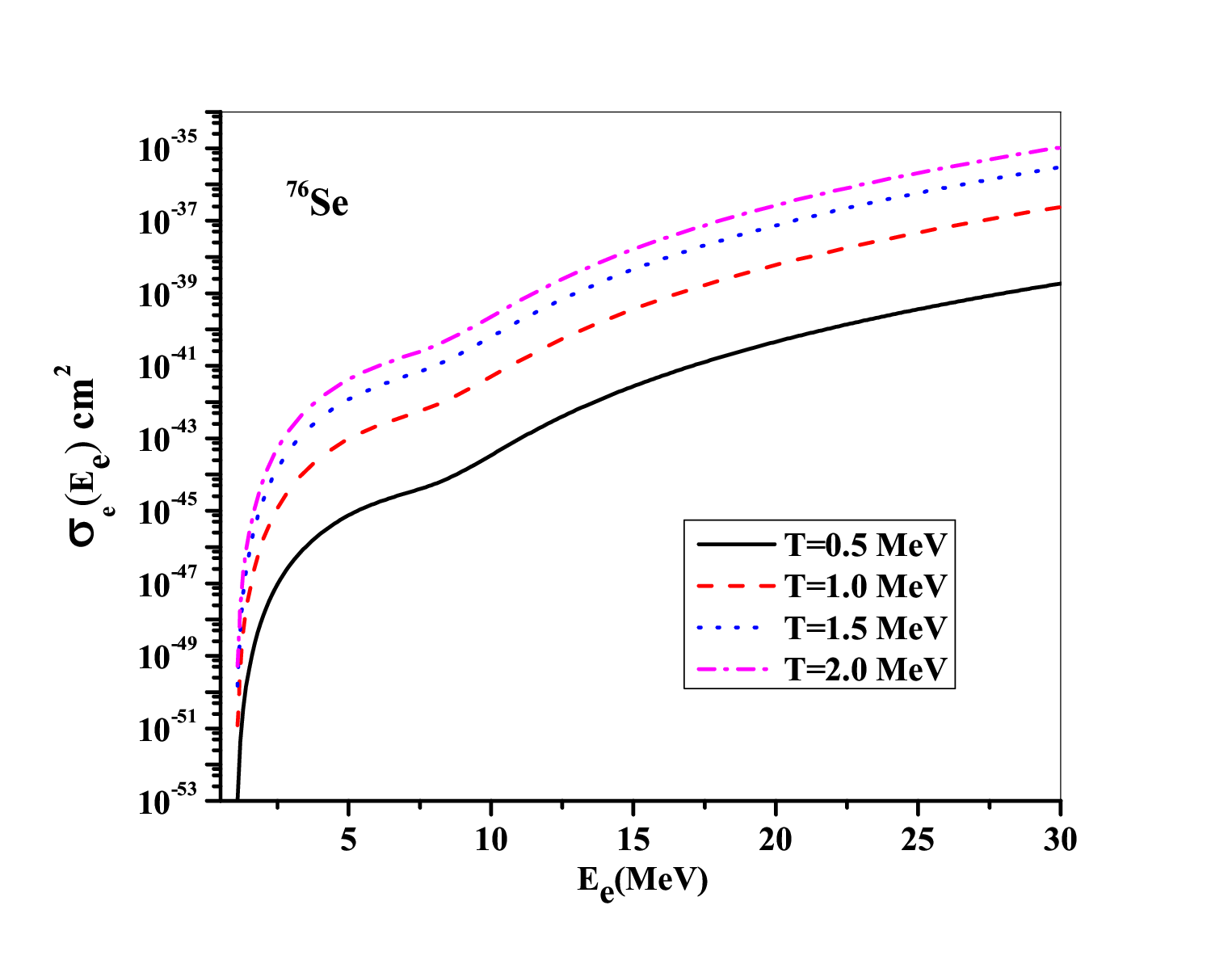}
\normalsize \caption{(Color online) The pn-QRPA calculated electron
capture cross sections as a function of the incident electron energy
($E_{e}$) at different stellar temperatures.}\label{cs}
\end{figure}

The calculated electron capture cross section (ECC) for $^{76}$Se
using pn-QRPA model is shown in the Fig~\ref{cs}. The calculated ECC
is shown as a function of temperature and incident electron energy.
The threshold electron energy to initiate the electron capture
process is dictated by the $Q$ value of the nuclear reaction.  The
ECC increases exponentially as incident electron energy increases up
to 4 MeV and later the slope decreases at high energies.  This
behavior maybe mapped to the trend of calculated GT distribution.
States having high multipoles contribute less so the ECC increases
smoothly as the incident electron energy increases. The temperature
dependance of ECC is also evident from Fig~\ref{cs}. As the stellar
temperature increases from 0.5 MeV to 1.0 MeV, one notes that the
ECC increases by roughly two orders of magnitude corresponding to a
significant thermal unblocking of the GT$^{+}$ channel. A further
increase in temperature does not produce matching increment in the
calculated ECC as majority of transitions are already thermally
unblocked.

\begin{figure}
\includegraphics[scale=0.5]{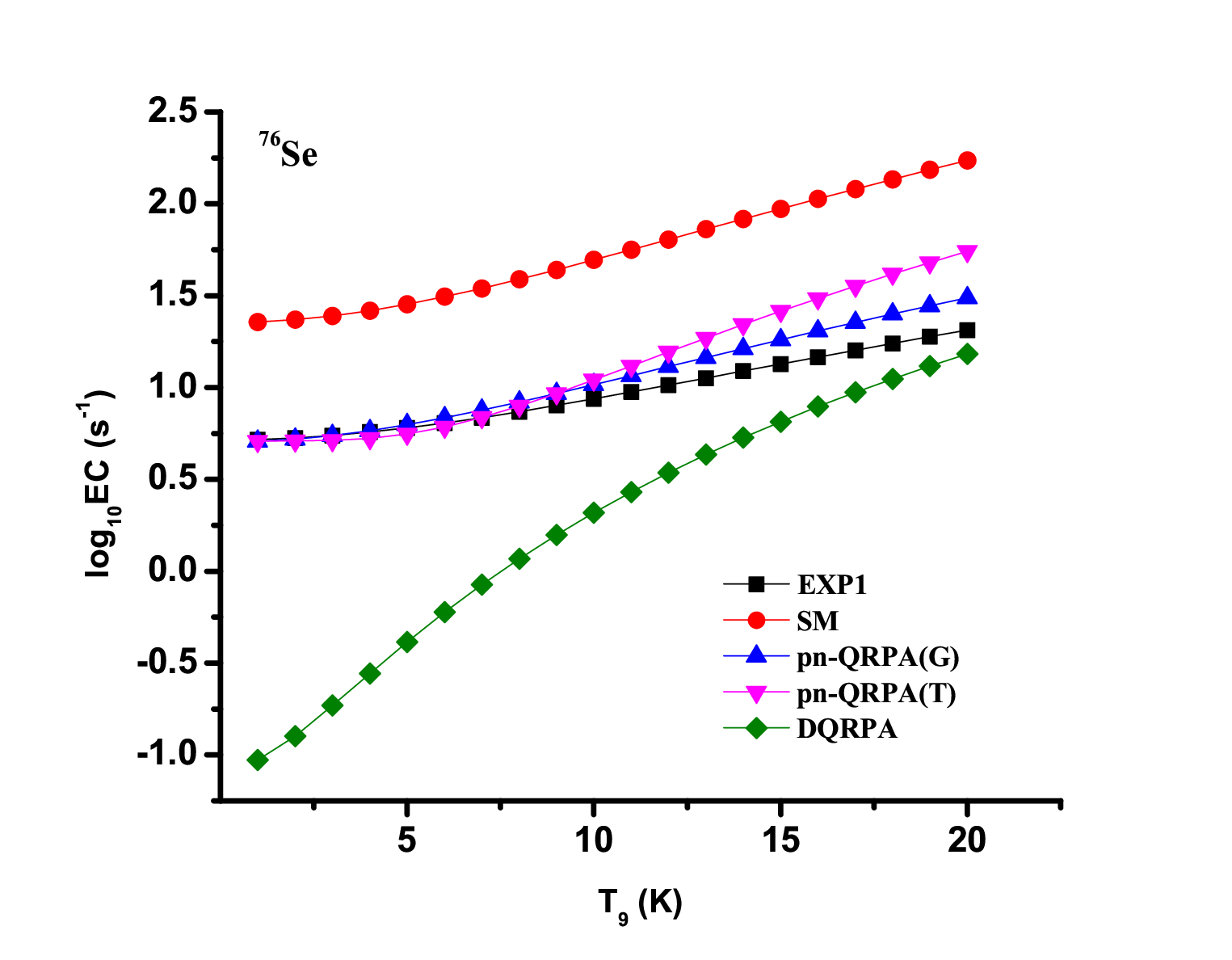}
\normalsize \caption{(Color online) Comparison of pn-QRPA calculated
electron capture (EC) rates using contribution from only ground
state (pn-QRPA(G)) and from all discrete excited states (pn-QRPA(T))
with EXP1 \cite{gre08}, shell model \cite{zhi11} and DQRPA
\cite{ha15} results, at fixed core density of $\rho$ = $10^{9.6}$
g$cm^{-3}$.}\label{aden}
\end{figure}

We next present the calculation of stellar electron capture rates on
$^{76}$Se. Fig.~\ref{aden} depicts the calculated EC rates as a
function of core temperature at a fixed stellar density of
$10^{9.6}$ gcm$^{-3}$. The ordinate shows calculated EC rates in
logarithmic scale having units of s$^{-1}$. Shown in Fig.~\ref{aden}
are the calculated EC rates using GT strength distributions from (i)
$^{76}$Se(d, $^{2}$He)$^{76}$As data \cite{gre08} (marked as EXP1),
(ii) shell model calculation \cite{zhi11} (marked as SM) and the
DQRPA calculation \cite{ha15}. Our calculated EC rates are
calculated using first only the ground state GT strength
distribution (marked as pn-QRPA(G)) and later from contribution of
all excited state GT strength distributions calculated
microscopically using our model and denoted by pn-QRPA(T) in
Fig.~\ref{aden}. In other words, apart from the pn-QRPA(T) data, all
EC rates were calculated using only the ground state GT strength
distribution  and may be compared mutually. Our code allows manual
insertion of energy levels and B(GT)$_{ij}$ strength values (see
Eq.~(\ref{gt})). We inserted the desired experimental and
theoretical data in our code to calculate the electron capture
rates.  We manually inserted $i$ = 1 in Eq.~(\ref{total rate}) for
EXP1, SM, pn-QRPA(G) and DQRPA calculated electron capture rates in
our code. Here we see that the DQRPA calculated EC rates are
smallest because of placement of GT centroid at much higher value
(see Table~\ref{centroid}). SM calculated EC rates are biggest. This
is because SM calculated a relatively big total strength and also
placed the GT centroid at relatively low energy in daughter
(Table~\ref{centroid}). At low stellar temperatures the pn-QRPA(G)
and pn-QRPA(T) rates are in decent comparison with experimental
data. This is because the slightly bigger centroid  (calculated in
the pn-QRPA model) effect is offset by the marginally bigger total
strength (calculated in the pn-QRPA model). It is also noted that
the pn-QRPA(G) rates are marginally bigger than the pn-QRPA(T) rates
till the core temperature reaches T$_{9}$ =10. This behavior is due
to the fact that for the pn-QRPA(G) rates, $P_{g.s.}$ = 1 (because
$i$ = 1). However for pn-QRPA(T) rates, $P_{g.s.} < $ 1 (there are
total of 180 discrete states calculated within the pn-QRPA model).
At low core temperatures where the excited state occupation
probability is not appreciable, the pn-QRPA(T) rates are slightly
smaller than pn-QRPA(G) rates because of marginally smaller
$P_{g.s.}$ value. As T$_{9} \sim$ 10, the partial rates from excited
states contribute significantly (higher corresponding $P_{i}$
values) and then total EC rates are bigger.

\begin{figure}
\includegraphics[scale=0.5]{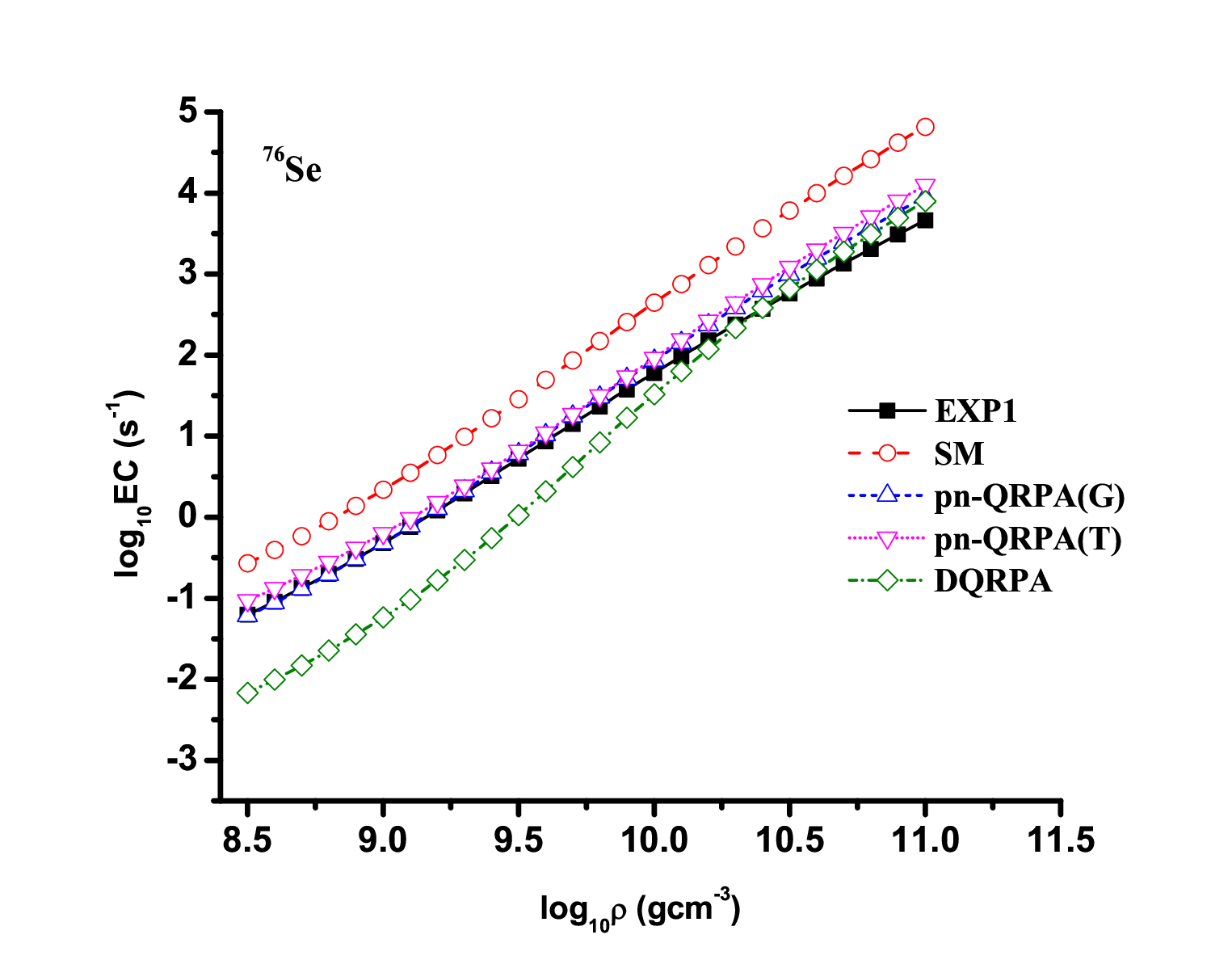}
\normalsize \caption{(Color online) Same as Fig.~\ref{aden} but at
fixed core temperature of T$_{9}$(K)=10.}\label{t9}
\end{figure}

Next we show the same EC calculations as a function of core density
and at a fixed temperature of T$_{9}$ =10 (Fig.~\ref{t9}). Here once
again the SM calculated EC rates are biggest. At high density DQRPA
rates are in better agreement with EXP1 EC rates. The pn-QRPA(G) and
pn-QRPA(T) rates are in decent comparison with experimental EC rates
specially at low density values. It is to be noted that at a fixed
temperature occupation probability of parent excited states remain
constant and the variation is due to varying nature of the
calculated phase space as the stellar core stiffens to higher
density. We did not show the pn-QRPA calculated excited state GT
strength distributions for space consideration and the same may be
requested as ASCII files from the corresponding author.

\begin{figure}
\includegraphics[scale=0.5]{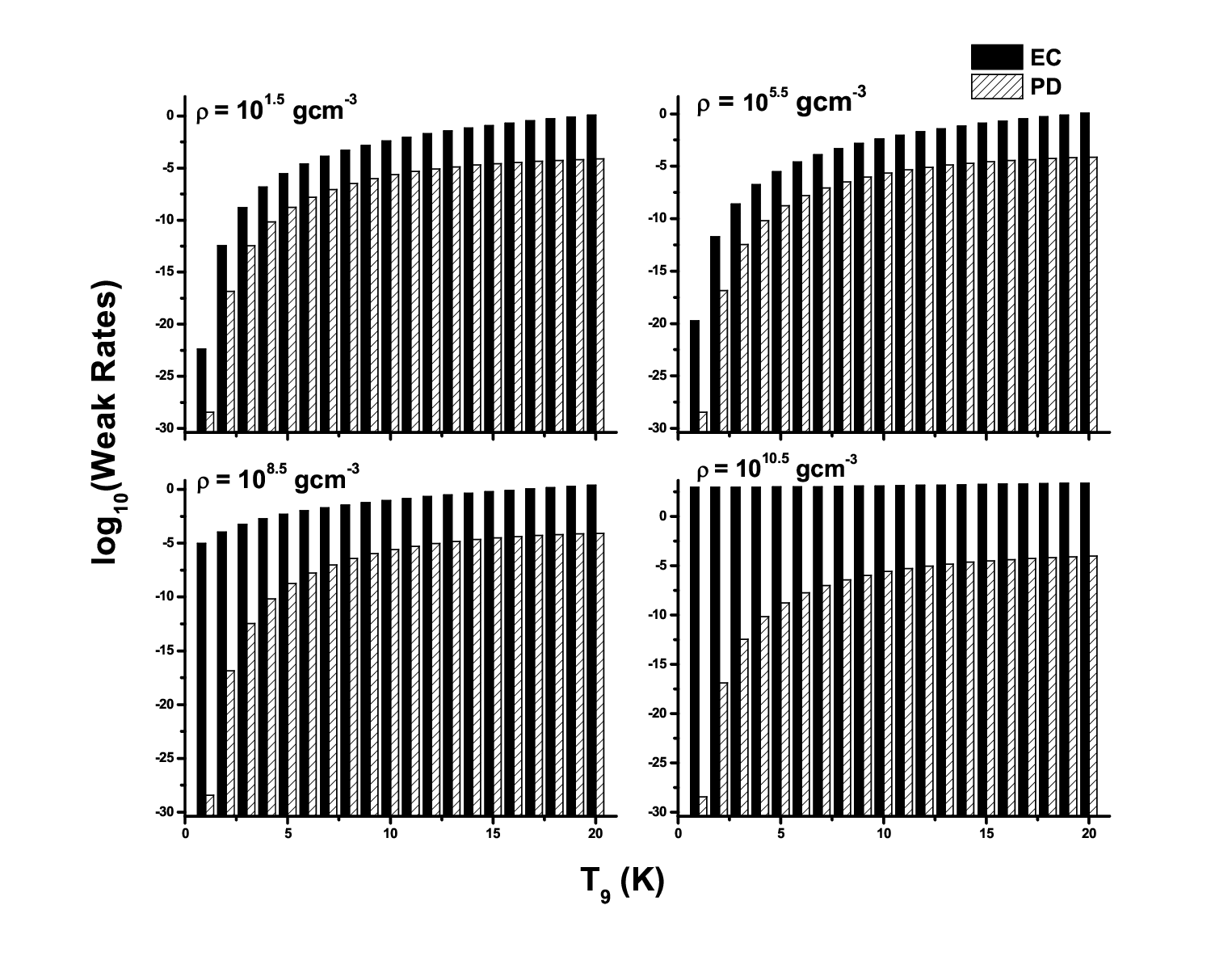}
\normalsize \caption{ The pn-QRPA calculated electron capture (EC)
and positron decay (PD) rates as a function of core density and
temperature.}\label{ecpd}
\end{figure}

Table~\ref{ratio} shows relative contribution of our calculated
$\beta^{+}$-decay and  electron capture rates  for $^{76}$Se as a
function of stellar temperature and density.  The calculated EC
rates  are  orders of magnitude bigger than the competing
$\beta^{+}$-decay rates at low stellar temperatures (T$_{9} \leq $
5). As the core temperature increases so does the contribution of
$\beta^{+}$-decay rates. In our calculation, it is assumed that at
high temperatures ($kT \ge$ 1 MeV), positrons appear via
electron-positron pair creation process. We performed a similar
study of relative contribution of $\beta^{+}$-decay rates on
$^{76}$Se in stellar matter on a wider density range.
Fig.~\ref{ecpd} shows the EC and positron decay (PD) contribution at
different T$_{9}$(K) temperatures and selected values of densities
10$^{1.5}$, 10$^{5.5}$, 10$^{8.5}$ and 10$^{10.5}$ gcm$^{-3}$. We
came to the conclusion that at high stellar temperatures and low
densities, the $\beta^{+}$-decay on $^{76}$Se may not be neglected
and needs to be taken into consideration along with EC rates for
better simulation results.

\begin{figure}
\includegraphics[scale=0.5]{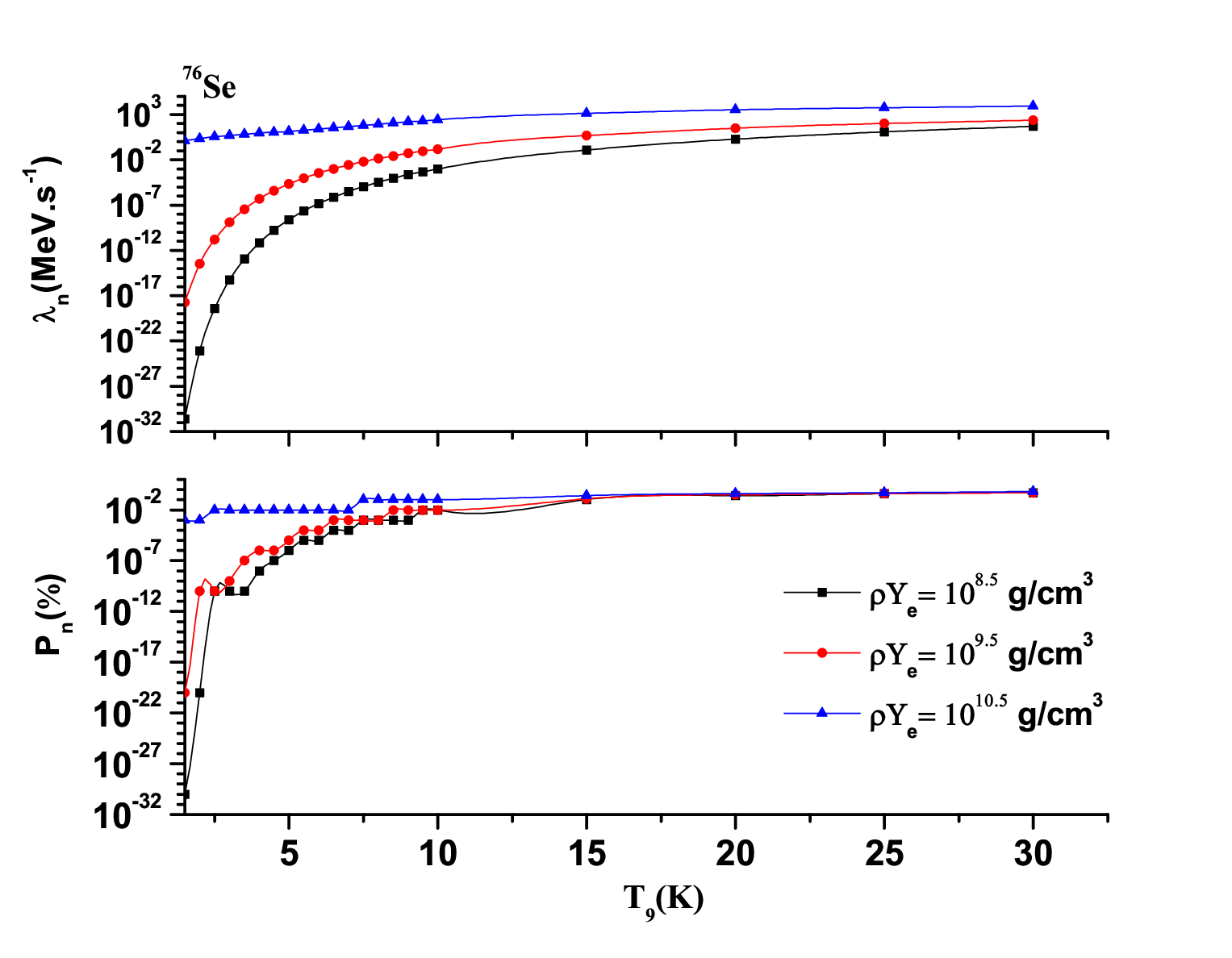}
\normalsize \caption{(Color online) Calculated $\beta$-delayed
neutron energy rate (top panel) and probability of $\beta$-delayed
neutron emissions (bottom panel) from $^{76}$As (daughter nucleus of
$^{76}$Se) as a function of stellar temperature and
density.}\label{pn76}
\end{figure}

The energy rate of $\beta$-delayed neutron emissions at selected
densities of 10$^{8.5}$, 10$^{9.5}$ and 10$^{10.5}$ gcm$^{-3}$ as a
function of core temperature is shown in upper panel of
Fig.~\ref{pn76}. The bottom panel depicts the probability of
$\beta$-delayed neutron emissions from $^{76}$As (daughter of
$^{76}$Se). The Q-value of the reaction is -2.96058 MeV, whereas
neutron separation energy of parent (daughter) nucleus used in our
calculation  is 11.15390 MeV (7.32861 MeV). These values were
obtained using data from \cite{aud12}. The energy rates of particle
emission (and corresponding probability) increases with increasing
density and temperature. At high stellar temperatures the rates tend
to merge into each other. This is because at high stellar
temperatures daughter states, with energy greater than neutron
separation energy, have a finite occupation probability. The
$\beta$-delayed neutron energy rate and probability of
$\beta$-delayed neutron emissions merge into one another as core
density stiffens from 10$^{8.5}$ to 10$^{10.5}$ gcm$^{-3}$ at high
T$_{9}$ values. Further at high stellar density of 10$^{10.5}$
gcm$^{-3}$, the Fermi energy is high enough even at low temperatures
and both $\beta$-delayed neutron energy rate and probability of
$\beta$-delayed neutron emissions are correspondingly higher. The
ASCII files of all calculated stellar rates are available and may be
requested from the corresponding author.

\section{Conclusion}

Recently, both IBM-1 and pn-QRPA models were applied to investigate
nuclear structure properties and weak rates of even-even nuclei with
the same proton and neutron numbers ~\cite{Nabi16,Nabi17}. In the
current work, we used two formalisms of the IBM-1 model for the
description of the nuclear structure properties of $^{76}$Se
nucleus. We first investigated the energy levels, $B(E2)$ values and
geometry of $^{76}$Se within the Hamiltonian in Eqs.~(\ref{ham}) and
(\ref{tri}). Later we used the pn-QRPA model to calculate GT
strength distributions and associated weak-interaction mediated
rates for $^{76}$Se. We also commented on unblocking of GT strength
in this paper.

$^{76}$Se exhibits typical spherical harmonic triplet ($4_{gs}^+$,
$2_{\gamma}^+$, $0_{\beta}^+$) at low energy spectrum. Its energy
ratio is $R_{4/2}=2.38$ and is bigger than the ratio of $U(5)$ which
hints that this nucleus can be located in between $U(5)$ and $O(6)$
symmetries (the spherical and the $\gamma$-~softness). Therefore, we
first chose the ECQF formalism in Eq.~(\ref{ham}) that allows to
move in between these symmetries. However, even--odd staggering

occurred in the $\gamma$-band levels and later we added the
\emph{cubic} term into the ECQF Hamiltonian to see the behavior of
the $\gamma$-~softness. Although the exact $\gamma$-~softness did
not appear, we obtained better results within the formalism of
IBM-1$^{*}$ because of the effect of the \emph{cubic} term. We
analyzed the $\gamma$-band and the signatures of $\gamma$-softness
by plotting the signature splitting S(J) to see its effect in
detail. Our calculation showed that this nucleus lies in between
spherical and $\gamma$-softness transitional path. The $cubic$ term
had a positive effect on the $\gamma$~band levels even though the
energy surfaces of IBM-1 calculations resulted in a spherical
minima.

In the later phase of our work we used the pn-QRPA model to
calculate GT strength distributions and stellar weak rates for
$^{76}$Se. We decided to use the deformation value extracted from E2
transition \cite{ram01} in calculation of our GT strength
distribution for $^{76}$Se. The choice was made due to the fact that
deformation parameter is one of the key parameters in pn-QRPA
calculation. As different theoretical models were suggesting varying
shapes for $^{76}$Se, we decided to use the value of $\beta_{2}$
extracted from experimental data in our pn-QRPA calculation. The
energy levels for $^{76}$Se, calculated by the IBM-1$^{*}$ model,
was subsequently incorporated in the GT and weak rate calculations
performed using the pn-QRPA model.

Our pn-QRPA calculation reconfirmed the  unblocking of GT strength
for $^{76}$Se (N $>$ 40 and Z $<$ 40, case). The chosen $\beta_{2}$
value resulted in a decent comparison of calculated ground state GT
strength distribution with experimental data. Later the same model
was used to calculate electron capture cross sections and weak rates
in stellar matter for $^{76}$Se. The main feature of our calculation
was a state-by-state calculation of all weak rates in a fully
microscopic fashion.  Our study implies that electron capture rate
is the dominant mode of decay for $^{76}$Se. However, at low stellar
densities and high stellar temperatures, $\beta^{+}$-decay on
$^{76}$Se should also be accounted for a better and realistic
interpretation of nuclear network calculations.

\section*{Acknowledgements}

J.-U. Nabi also wishes to acknowledge the support provided by the
Higher Education Commission (Pakistan) through the HEC Projects No.
20-3099 and 20-5557.

\newpage
%FIGURES

\newpage
\begin{table}
\caption{The fitted parameters of the IBM-1 Hamiltonian
(Eq.~(\ref{ham})) in units of keV.  $\overline{\chi}$ is
dimensionless, $N$ is the boson number, $v_{3}$ is the cubic
interaction and $\sigma$ is $rms$ value.} \label{par} \centering
\begin{tabular}{cccccccc}
\hline
$^{76}$Se & $N$ &$\epsilon_{d}$&$a_{1}$&$a_{2}$&$v_{3}$&$\overline{\chi}$&$\sigma$\\
\hline
IBM-1~&~~7~~~&~~524.4~~~&~~~14.5~~~&~~~5.7~~~&~~~ - ~~~&~~-0.95~~~&~~94~\\
IBM-1$^*$&~~7~~~&~~624.9~~~&~~~11.6~~~&~~-8.7~~~&~~~ -81.6 ~~~&~~-0.95~~~&~~68~\\
\hline
\end{tabular}
\end{table}

\begin{table}
\caption{Measured and calculated B(E2) values in units of $10^{-2}$
$e^{2}b^{2}$ for  $^{76}$Se.} \label{be2} \centering
\begin{tabular}{@{}lllll}
\hline
~~~$J^\pi_i\rightarrow J^\pi_f$~~~~&~~~~~~EXP~~~~~~&~~~~~~IBM-1~~~~~~&~~~~~~IBM-1$^*$~~\\
\hline
~~~$2^+_{gs}\rightarrow0^+_{gs}$~~~~&~~~~~~8.41 (\emph{0.28})~~~~~~&~~~~~~5.82~~~~~~&~~~~~~7.82~~~~\\
~~~$4^+_{gs}\rightarrow2^+_{gs}$~~~~&~~~~~~13.58 (\emph{0.55})~~~~~~&~~~~~~9.97~~~~~~&~~~~~~13.41~~~~\\
~~~$6^+_{gs}\rightarrow4^+_{gs}$~~~~&~~~~~~13.00 (\emph{2.21})~~~~~~&~~~~~~12.43~~~~~~&~~~~~~16.65~~~~\\
~~~$8^+_{gs}\rightarrow6^+_{gs}$~~~~&~~~~~~15.68 (\emph{4.69})~~~~~~&~~~~~~13.17~~~~~~&~~~~~~17.57~~~~\\
~~~$10^+_{gs}\rightarrow8^+_{gs}$~~~~&~~~~~~9.94 (\emph{3.03})~~~~~~&~~~~~~12.21~~~~~~&~~~~~~16.26~~~~\\
~~~$2^+_{\gamma}\rightarrow0^+_{gs}$~~~~&~~~~~~0.23 (\emph{0.03})~~~~~~&~~~~~~0.005~~~~~~&~~~~~~0.02~~~~\\
~~~$3^+_{\gamma}\rightarrow2^+_{gs}$~~~~&~~~~~~0.36 (\emph{0.22})~~~~~~&~~~~~~0.008~~~~~~&~~~~~~0.03~~~~\\
~~~$4^+_{\gamma}\rightarrow4^+_{gs}$~~~~&~~~~~~4.21 (\emph{1.93})~~~~~~&~~~~~~6.49~~~~~~&~~~~~~6.33~~~~\\
~~~$4^+_{\gamma}\rightarrow2^+_{\gamma}$~~~~&~~~~~~5.55 (\emph{1.93})~~~~~~&~~~~~~6.75~~~~~~&~~~~~~8.09~~~~\\
~~~$5^+_{\gamma}\rightarrow3^+_{\gamma}$~~~~&~~~~~~12.81 (\emph{6.34})~~~~~~&~~~~~~7.36~~~~~~&~~~~~~8.58~~~~\\
~~~$5^+_{\gamma}\rightarrow4^+_{gs}$~~~~&~~~~~~0.90 (\emph{0.47})~~~~~~&~~~~~~0.008~~~~~~&~~~~~~0.02~~~~\\
~~~$6^+_{\gamma}\rightarrow4^+_{\gamma}$~~~~&~~~~~~5.55 (\emph{4.96})~~~~~~&~~~~~~9.34~~~~~~&~~~~~~11.31~~~~\\
~~~$7^+_{\gamma}\rightarrow5^+_{\gamma}$~~~~&~~~~~~7.65 (\emph{5.79})~~~~~~&~~~~~~8.99~~~~~~&~~~~~~10.6~~~~\\
~~~$0^+_{\beta}\rightarrow2^+_{gs}$~~~~&~~~~~~8.99 (\emph{6.07})~~~~~~&~~~~~~10.48~~~~~~&~~~~~~12.51~~~~\\
\hline
\end{tabular}
\end{table}

\begin{table}
\caption{Deformation value of $^{76}$Se using RMF model
\cite{lal99}, IBM-1 (this work) and that extracted from the E2
transition \cite{ram01}.}\label{dv}
\begin{tabular}{cccc}
\hline
~~~~~~~~~&~~~~~~~~RMF~~~~~~~&~~~~~~~~~IBM-1~~~~~~&~~~~~~~~~E2 transition~~~~~\\
\hline\noalign{\smallskip}
~~~~~$\beta_{2}$~~~~&~~~~~~~~~-0.244~~~~~~&~~~~~~~~~0.000~~~~~~&~~~~~~~~~0.309~~~~~~~\\
\noalign{\smallskip}\hline
\end{tabular}
\vspace*{6 cm}
\end{table}

%\begin{table}
%\caption{Deformation value of $^{76}$Se using RMF model
%\cite{lal99}, IBM (this work) and that extracted from the E2
%transition \cite{ram01}.}\label{dv}
%\begin{tabular}{cccc}
%\hline\noalign{\smallskip}
%$\beta_{2}$ (RMF)& -0.244  \\
%$\beta_{2}$ (IBM)& 0.000  \\
%$\beta_{2}$ $(E_{2})$ & 0.309  \\
%\noalign{\smallskip}\hline
%\end{tabular}
%\vspace*{6 cm}
%\end{table}

\begin{table}
\centering \caption{Total B(GT) strengths, centroids and cut-off
energy values for  $^{76}$Se in electron capture
direction.}\label{centroid}
    \begin{tabular}{c|c|c|c}

           Model &$\sum$ B(GT$_{+}$) & $\bar{E}_{+}$ (MeV) & Cut-off energy(MeV)\\
\hline

           EXP1 \cite{gre08} & 1.00 & 2.50 & 10.00  \\
           EXP2 \cite{hel97} & 0.98 & 2.47 & 6.00 \\
           EXP3 \cite{hel97} & 0.35 & 1.16 & 6.00\\
           SM \cite{zhi11} & 1.46 & 2.79 & 9.80 \\
           pn-QRPA (this work) & 1.41 & 3.09 & 12.00\\
           DQRPA  \cite{ha15} & 2.04 & 7.48 & 25.00 \\

           \end{tabular}
\end{table}

\begin{table}
\centering  \caption{Ratio of pn-QRPA calculated electron capture
(EC)  to $\beta^{+}$ decay rates as a function of stellar density
(in units of g/cm$^{3}$) and temperature (in units of 10$^{9}$
K).}\label{ratio}
    \begin{tabular}{c|c|c|c|c}

           $\rho$$\it Y_{e}$  & \multicolumn{3}{c}{$R(EC/\beta^{+})$}\\
\cline{2-5}  &T$_{9}$=1 & T$_{9}$=5 & T$_{9}$=10 & T$_{9}$=20 \\
\hline

           10$^{8.5}$ & 2.60E+23 & 2.63E+06 & 3.62E+04 & 2.98E+04 \\
           10$^{9.5}$ & 6.64E+28 & 1.60E+09 & 2.52E+06 & 4.00E+05 \\
           10$^{10.5}$ & 2.51E+31 & 5.38E+11 & 4.65E+08 & 2.50E+07 \\
           \hline

           \end{tabular}
\end{table}

\section*{References}
\begin{thebibliography}{99}
\bibitem{bet79} H. A. Bethe, G. E. Brown, J. Applegate, J. M. Lattimer, Nucl. Phy. A, 324 (1979) 487.
\bibitem{bet90} H. A. Bethe, Rev. Mod. Phy. 62 (1990) 801.
\bibitem{Niu11} Y. F. Niu, N. Paar, D. Vretenar, J. Meng, Phy. Rev. C 83 (2011) 045807.
\bibitem{paa09} N. Paar, G. Colo, E. Khan, D. Vretenar, Phys. Rev.
C 80 (2009) 055801.
\bibitem{fan12} A. F. Fantina, E. Khan, G. Col\`o, N. Paar, D. Vretenar, Phys. Rev. C 86 (2012) 035805.

\bibitem{Moe97} P. M\"oller, J. R. Nix, K.-L. Kratz, At. Data. Nucl. Data. Tables 66 (1997) 131.
\bibitem{Hir93} M. Hirsch, A. Staudt, H. V. Klapdor-Kleingrothaus, At. Data. Nucl. Data. Tables 51 (1992) 243.
\bibitem{Bor06} I. N. Borzov, Nucl. Phy. A 777 (2006) 645.
\bibitem{Mar99} G. Martinez-Pinedo, K. Langanke, Phy. Rev. Lett. 83 (1999) 4502.
\bibitem{Lan95} K. Langanke, D. J. Dean, P. B. Radha, Y. Alhassid, S. E. Koonin, Phy. Rev. C 52 (1995) 718.
\bibitem{Mut91}K. Muto, E. Bender, T. Oda, Phys. Rev. C 43 (1991) 1487.
\bibitem{Iachello87} F. Iachello, A. Arima, The Interacting Boson Model, Cambridge University Press, Cambridge, 1987.
\bibitem{Kaup83} U. Kaup, C. M\"onkemeyer, P. Von~Brentano, Z. Phy. A 310 (1983) 129.
\bibitem{Radhi86} F. S. Radhi, N. M. Stewart, Z. Phy. A 356 (1986) 145.
\bibitem{Speidel98} K. H. Speidel, Phy. Rev. C 57 (1998) 2181.
\bibitem{Turkan06} N. T\"urkan, D. Olgun, I. Uluer, Cent. Eur. J. Phy 4 (2006) 124.
\bibitem{Boyukata08} M. B\"oy\"ukata, I. Uluer, Cent. Eur. J. Phy 6 (2008) 518.
\bibitem{Lipas85} P. O. Lipas, P. Toivonen, D. D. Warner, Phy. Lett. B 155 (1985) 295.
\bibitem{Heyde84} K. Heyde, P. V. Isacker, M. Waroquier, J. Moreau, Phy. Rev. C 29 (1984) 1420.
\bibitem{Hal67} J. A. Halbleib, R. A. Sorenson, Nucl. Phy. A 98 (1967) 542.
\bibitem{Bri55} D. Brink, D. Phil. Thesis, Oxford University, Unpublished 1955; P. Axel, Phys. Rev. 126 (1962) 671.
\bibitem{ful82} G. M. Fuller, W. A. Fowler, M. J. Newman, Astrophys. J 252 (1982) 715; G. M. Fuller, W. A. Fowler, M. J. Newman, Astrophys. J. 293 (1985) 1.
\bibitem{ful82a} G. M. Fuller, Astrophys. J. 252 (1982) 741.
\bibitem{nab99} J. -U. Nabi, H. V. Klapdor-Kleingrothaus, Eur. Phys. J. A 5 (1999) 337.
\bibitem{nab99a} J. -U. Nabi, H. V. Klapdor-Kleingrothaus, At. Data. Nucl. Data. Tables 71 (1999) 149.
\bibitem{nab04} J. -U. Nabi, H. V. Klapdor-Kleingrothaus, At. Data. Nuc. Data. Tables 88 (2004) 237.
\bibitem{lan00} K. Langanke, G. Martinez-Pinedo, Nucl. Phy. A 673 (2000) 481.
\bibitem{Nabi16} J. -U. Nabi, M. B\"oy\"ukata, Nucl. Phy. A 947 (2016) 182.
\bibitem{Nabi17} J. -U. Nabi, M. B\"oy\"ukata, Astrophys Space Sci. 362 (2017) 9.
\bibitem{Lan03} K. Langanke \textit{et al.}, Phys. Rev. Lett. 90 (2003) 241102.
\bibitem{Hix03} R. W. Hix \textit{ et al.}, Phys. Rev. Lett. 91 (2003) 210102.
%\bibitem{lan01} K. Langanke, E. Kolbe, D. J. Dean, Phys. Rev. C 63 (2001)
%032801.
\bibitem{coo84} J. Cooperstein, J. Wambach, Nucl. Phys. A 420
(1984) 591.
\bibitem{gre08} E. W. Grewe, C. Baumer, H. Dohmann, D. Frekers, M. N. Harakeh, S. Hollstein, H. Johansson, L. Popescu, S. Rakers, D. Savran, H. Simon, J. H. Thies, A. M. Berg, H. J. Wortche, A. Zilges, Phy. Rev. C 78 (2008) 044301.
\bibitem{hel97} R. L. Helmer, M. A. Punyasena, R. Abegg, W. P. Alford, A. Celler, S. El-Kateb, J. Engel, D. Frekers, R. S. Henderson, K. P. Jackson, S. Long, C. A. Miller, W. C. Olsen, B. M. Spicer, A. Trudel, M. C. Vetterli, Phy. Rev. C 55 (1997) 6.
\bibitem{sar03} P. Sarriguren, E. M. D. Guerra, L. Pacearescu, A. Faessler, F. Simkovic, A. A. Raduta, Phy. Rev. C 67 (2003) 044313.
\bibitem{zhi11} Q. Zhi, Y. Yu, Q. Zheng, Chi. Phys. C 35 (2011) 1022--1025.

\bibitem{ha15} E. Ha, M. K. Cheoun, Nucl. Phy. A 934 (2015) 73.
\bibitem{Iachello06} F. Iachello, Lie Algebras and Applications, Springer-Verlag, Berlin, 2006.
\bibitem{Warner82} D. D. Warner, R. F. Casten, Phy. Rev. Lett 48 (1982) 1385.
\bibitem{Warner83} D.D. Warner, R.F. Casten, Phys. Rev. C 28 (1983) 1798.
\bibitem{NNDC16} National Nuclear Data Center (NNDC), http://www.nndc.bnl.gov/, 2016.
\bibitem{Castanos84} O. Casta$\tilde{}$nos, A. Frank, P. Van Isacker, Phys. Rev. Lett. 52 (1984) 263.
\bibitem{Casten85a} R. F. Casten, P. Von Brentano, K. Heyde, P. Van Isacker, J. Jolie, Nucl. Phys. A 439 (1985) 289.
\bibitem{Casten85b} R. F. Casten, P. Von Brentano, Phys. Lett. B 152 (1985) 22.
\bibitem{Zamfir91} N. V. Zamfir, R. F. Casten, Phy. Lett. B 260 (1991) 265.
\bibitem{Stefanescu07} I. Stefanescu, A. Gelbergb, J. Jolie, P. Van Isacker, P. Von Brentano, Y. X. Luod, S. J. Zhuf, J. O. Rasmussene, J. H. Hamilton, A. V. Ramayya, X. L. Che, Nucl. Phys. A 789 (2007) 125.
\bibitem{Sorgunlu08} B. Sorgunlu, P. V. Isacker, Nucl. Phy. A 808 (2008) 27.
\bibitem{Fortunato11} L. Fortunato, C. E. Alonso, J. M. Arias, J. E. Garc$\acute{\i}$a-Ramos, A. Vitturi, Phys. Rev. C 84 (2011) 014326.
\bibitem{Dieperink80} A. E. L. Dieperink, O. Scholten and F. Iachello, Phys. Rev. Lett. 44 (1980) 1747.; Nucl. Phys. A 346 (1980) 125.
%\bibitem{Dieperink80b} A. E. L. Dieperink, O. Scholten, Nucl. Phys. A 346 (1980) 125.
\bibitem{Ginocchio80} J. N. Ginocchio, M. W. Kirson, Phys. Rev. Lett. 44 (1980) 1744.; Nucl. Phys. A 350 (1980) 31.
%\bibitem{Ginocchio80b} J. N. Ginocchio, M. W. Kirson, Nucl. Phys. A 350 (1980) 31.
\bibitem{Isacker81} P. Van~Isacker and J.-Q. Chen, Phys. Rev. C 24 (1981) 684.
\bibitem{Bohr98} A. Bohr and B. R. Mottelson, Nuclear Structure. Volume 2: Nuclear Deformation, World Scientific Publishing, 1998.
\bibitem{Nil55} S. G. Nilsson, Mat. Fys. Medd. Dan. Vid. Selsk 29  (1955) 16.
\bibitem{aud12} G. Audi, M. Wang, A. H. Wapstra, F. G. Kondev, M. MacCormick, X. Xu, B. Pfeiffer, Chin. Phy. C 36 () 1287, 2012; M. Wang, G. Audi, A. H. Wapstra, F. G. Kondev, M. MacCormick, X. Xu, B. Pfeiffer B, Chin. Phy. C 36 (2012) 1603.
\bibitem{Gov71} N. B. Gove, M. J. Martin, Nucl. Data Tables 10 (1971) 205.
\bibitem{jok02} A. Jokinen, A. Nieminen, J. \"{A}yst\"{o}, R. Borcea, E. Caurier, P. Dendooven, M. Gierlik, M. G\'{o}rska, H. Grawe, M. Hellstr\"{o}m, M. Karny, Z. Janas, R. Kirchner, M. L. Commara, G. Mart\'{i}nez-Pinedo, P. Mayet, H. Penttil\"{a}, A. Plochocki, M. Rejmund, E. Roeckl, M. Sawicka, C. Schlegel, K. Schmidt, R. Schwengner, Eur. Phys. J. A 3 (2002) 1.

\bibitem{nak10} K. Nakamura, Particle Data Group, J. Phys. G. Nucl. Part. Phys. 37(7A) (2010) 075021.
\bibitem{nab15b} S. Cakmak, J.-U. Nabi, T. Babacan, I. Maras, Adv. Space Research 55 (2015) 440.
\bibitem{mut92} K. Muto, E. Bender, T. Oda, H. V. Klapdor-Kleingrothaus, Z. Phy. A 341 (1992) 407.
\bibitem{div13} P. C. Divari, Adv. High. Energy. Phy 2013 (2013) 11.
\bibitem{Goo80} C. D. Goodman, C. A. Goulding, M. B. Greenfield, J. Rapaport, D. E. Bainum, C. C. Foster, W. G. Love, F. Petrovich, Phy. Rev. Lett 44 (1980) 1755.

\bibitem{Nab16} J.-U. Nabi, A. N. Tawfik, N. Ezzelarab, A. A. Khan, Phys. Scripta 91 (2016) 055301.
\bibitem{Casten90} R. F.  Casten, Nuclear  Structure  From  a  Simple  Perspective, Oxford  U.P., Oxford, 1990.
\bibitem{Boyukata10} M. B\"oy\"ukata, P. Van~Isacker, \.{I}.\ Uluer, J. Phy. G: Nucl. Part. Phy 37 (2010) 105102.
\bibitem{lal99} G. A. Lalazissiz, S. Raman, P. Ring, At. Data Nucl. Data Tables 71 (1999) 1.
\bibitem{ram01} S. Raman, C. W. Nestor, Jr., P. Tikkanen, At. Data Nucl. Data Tables 78 (2001) 1-128.
\end{thebibliography}
\end{document}